# Growth of hydroxyapatite in a biocompatible mesoporous ordered silica

A. Díaz [a,*], T. López [b,c], J. Manjarrez [c], E. Basaldella [d],
J.M. Martínez-Blanes [a], J.A. Odriozola [a]

[a] *Departamento de Química Inorgánica e Instituto de Ciencia de Materiales de Sevilla (Centro Mixto CSIC-Universidad de Sevilla), Centro de Investigaciones Científicas Isla de la Cartuja, Avda, Américo Vespucio no. 49, 41092 Sevilla, Spain*
[b] *Universidad Autónoma Metropolitana Iztapalapa, P.O. Box 55-534, México, DF 09340, México*
[c] *Instituto Nacional de Neurología y Neurocirugía "Manuel Velasco Suárez", Insurgentes Sur 3877, CP 14269 México, DF, México*
[d] *CIC-CINDECA, Universidad Nacional de La Plata, calle 47, No. 257 (1900), La Plata, Argentina*

## Abstract

A novel biomaterial (HA-SBA-15) has been developed based on the growth of calcium phosphate hydroxyapatite (HA) nanoparticles within an organized silica structure (SBA-15). Characterization of the material was carried out using a combination of X-ray diffraction, X-ray fluorescence, transmission electron microscopy, $N_2$ adsorption–desorption isotherms and nuclear magnetic resonance. Transmission electron microscopy observations and $N_2$ porosimetry revealed the crystallization of hydroxyapatite nanoparticles inside the mesopore cavities of the silica structure. Specific surface areas of 760 $m^2\,g^{-1}$ and 260 $m^2\,g^{-1}$ were measured for the SBA-15 and the HA-SBA-15 material, respectively. The hydroxyl groups present in the silica nanostructure surface have brought about cationic defects in the silicium sites, mainly with those of tetrahedral symmetry, and promoted the formation of siloxanes. $^{29}Si$ MAS-NMR analysis shows a significant reduction of the silanol groups concentration with HA growing within the base (SBA-15) material. Studies and brain tissue biocompatibility tests were carried out. Histopathological studies on the SBA-15 implant material showed no changes to the tissue nearby. The results confirmed the synthesis of a silica-based composite containing HA nanoparticles with the potential for biomedical applications.

*Keywords:* Nanostructured materials; Mesoporous silica; Biocompatibility; Hydroxyapatite

## 1. Introduction

Skeletal deficiencies, especially resulting from trauma, tumors or abnormal development, are common, and are usually treated by surgical intervention and grafting to restore mechanical function and reconstruct the operation area. The steadily increasing number of skeletal deficiencies requires bone replacements and effective ways to enhance bone formation [1]. This problem has led scientists to find a biomaterial with grafting properties which possess biocompatibility, functionality and simplicity and is also economically viable [2,3]. Bone tissue engineering may potentially provide alternative solutions that possess better fixation and mechanical properties than those used currently. Bone tissue engineering is the use of a scaffolding material either to induce formation of bone from the surrounding tissue or to act as a carrier or template for implanted bone cells or other agents. Materials used as bone tissue-engineered scaffolds may be injectable or rigid, the latter requiring an operative implantation procedure [4,5]. Composites obtained by nucleation and growth of inorganic phases inside the pores of bioceramics are a good alternative. Hydroxyapatite (HA) can be considered as a leading biomaterial satisfying these conditions. It is the main mineral component of bone and teeth, having the chemical formula of $Ca_{10}(PO_4)_6(OH)_2$.

* Corresponding author. Tel.: +34 954489500x9218; fax: +34 954460665.
*E-mail address:* aranzazu@icmse.csic.es (A. Díaz).

For bone and soft tissues, interfacial bonding with biomaterials occurs because of the biological equivalence of the inorganic portion of the bone and the growing hydroxyl carbonate apatite layer (HCA) on the bioactive implant. This interfacial bonding has been found to be the result of a complex dissolution, polycondensation and readsorption process which involves inorganic and organic ad-species and therefore can be promoted with increasing surface area. Recent molecular biology studies have shown that critical concentrations of soluble Ca and Si ions released from the bioactive materials control the cell cycle of osteoprogenitor cells [6]. The hierarchical organization of bone involves organic and inorganic nanophases. Hartgerink et al. [7] claim that one way to accomplish this hierarchical organization in an artificial system is to prepare an organic nanophase designed to exert control over crystal nucleation and growth of the inorganic component. Bioinspired morphogenesis of different inorganic phases including phosphate and apatite crystals is currently under intensive research [8–10].

HA has been used as bone graft but it has poor rates of reactivity and integration with existing bone when compared with bioactive glasses and glass–ceramics. The incorporation of Si atoms into HA increases the dissolution processes, enhancing the rate of incorporation of the graft though the precipitation of biological apatite [11].

Materials in the ternary system $SiO_2$–$CaO$–$P_2O_5$ obtained by a sol–gel process have demonstrated excellent bioactivity. From the early stages of biomaterials research it was clear that a high surface area of the bioactive material plays a substantial role in increasing bioactivity [12,13]. Both a negative surface charge and a porous substrate have been reported to be required for HCA formation [14,15]. An HCA layer can be formed on a porous pure silica gel in a solution containing $Ca^{2+}$ and $HPO_4^{2-}$ ions. The layer formation is enhanced by the presence of pores of between 2 and 50 nm diameter in sol–gel derived glasses, which increased the surface area/volume ratio compared to that of melt derived glasses [15].

An alternative to the use of organic templates for the hierarchical control of crystal growth could be the use of mesoporous ordered inorganic materials. Mesoporous SBA-15 type silica's have surface areas in the range of 600–1000 $m^2 g^{-1}$ and may accomplish the hierarchical requirements for growing apatite crystals resulting in a high surface area organized $SiO_2$–$CaO$–$P_2O_5$ system.

This paper reports the synthesis and the physicochemical characterization of a HA–silica composite biomaterial which has the potential to be used in the development of bone-fixation devices and could also be considered as a porous matrix for performing in situ controlled delivery of chemotherapy drugs.

## 2. Materials and methods

### 2.1. Synthesis

The synthesis of the material has been carried out using a two-step procedure which is detailed in the following two sections:

#### 2.1.1. Preparation of the Ca-doped silica matrix

The first step (low pH step) consisted of the preparation of a calcium doped silica matrix which was obtained by means of a variation of the methodology described in Ref. [16]. Pluronic® 123 block copolymer ($EO_{20}PO_{70}EO_{20}$; molecular weight 5750 g/mol) donated from the BASF Corporation was used as a template. The block copolymer was dissolved in an aqueous solution containing HCl (Panreac, PA-ACS-ISO 37%) and $CaCl_2 \cdot 2H_2O$ (Alfa-Aesar 99%) at a pH of ≈1 and 313 K. Tetraethyl orthosilicate (TEOS) (Alfa-Aesar 99%) was used as the silica source. The mixture was stirred for 24 h and subjected to a hydrothermal treatment at 353 K for 24 h. The product was dried in an oven at 353 K.

Using similar synthesis conditions a base mesoporous silica (SBA-15), without $CaCl_2 \cdot 2H_2O$ addition, and heat-treated in air at 773 K, was also synthesized as reference material.

#### 2.1.2. HA crystallization inside of the Ca-doped matrix

In the second step (high pH step) the Ca-doped silica was dropped into a $(NH_4)_2HPO_4$ (Merck P99%) solution and was submitted to a second hydrothermal treatment at 353 K for 24 h. The powder was then washed and heat-treated in air at 773 K to obtain the final (HA-SBA-15) material.

### 2.2. Materials characterization

#### 2.2.1. Transmission electron microscopy

Micrographs were recorded using a Philips CM200 transmission electron microscope (TEM) with a $LaB_6$ filament as the electron source, operated at 200 kV. Material samples were mounted on a microgrid carbon polymer, supported on a copper grid, by placing a few droplets of a suspension of the sample in water followed by drying at ambient conditions.

#### 2.2.2. X-ray diffraction analysis

X-ray diffraction (XRD) powder analysis was performed with a Philips X'Pert diffractometer using Cu-$K_a$ radiation (k = 0.15418 nm). The diffractometer was operated at 40 kV and 40 mA. Small-angle X-ray diffraction (SXRD) measurements were carried out using a step size of 0.002°, a 20 s exposure time and a position sensitive detector (PSD) length of 0.5170 (2h). Wide-angle X-ray diffraction (WXRD) patterns were recorded using a step size of 0.05°, a 100 s exposure time and a PSD length of 2.1180 (2h).

### 2.2.3. X-ray fluorescence analysis

X-ray fluorescence quantitative measurements were carried out using a sequential Siemens SRS3000 spectrometer with an Rh-anode X-ray tube with front window (75 lm). Characteristic $K\alpha$ lines were registered by a flow counter detector. Si and P line measurements were taken using 120 s, 30 kV, 100 mA and pentaerythrite (PET) crystal analyser. Ca line was obtained using 120 s, 50 kV, 60 mA and LiF 200 crystal analyser.

### 2.2.4. $N_2$ adsorption–desorption measurements

$N_2$ adsorption–desorption isotherms were collected on a Micromeritics ASAP2020 gas adsorption analyzer at 77 K, after degassing the samples at 523 K overnight on a vacuum line. The surface areas were calculated by the BET method and the pore size distributions were calculated from the desorption branch of the isotherm by the BJH method.

### 2.2.5. Nuclear magnetic resonance

$^{29}Si$ magic-angle-spinning (MAS) nuclear magnetic resonance (NMR) spectra were carried out with a Bruker DRX400 AVANCE (9.4 T) at a resonance frequency of 79.9 MHz. Powdered materials were spun at 11 kHz in a 4 mm outer diameter zirconia rotor, with TMS used as the reference and with a 2.75 ls pulse length and 600 s delay.

### 2.2.6. Biocompatibility tests

For the biological study six adult male Wistar rats (250–270 g) were used. One 1 · 1.5 mm SBA-15 silica cylinder and 1.3 mg weight (q = 1.1 g cm$^{-3}$) implant was prepared using uni-axial pressing and was surgically implanted though a stainless steel canula into the basolateral amygdale. After surgery the animals were allowed to recover in their home cage with food and water. At the end of the experiment, brains were removed and post-fixed in 3.7% formalin. The implantation sites were verified in coronal slices stained with hematoxylin eosin. Samples were evaluated using optical microscopy. The experiments were performed under the guidelines of the Mexican Law of Animal Protection.

## 3. Results and discussion

Fig. 1a and b show TEM micrographs of the SBA-15 and the HA-SBA-15 materials. Fig. 1b shows that the two-dimensional (2D)-hexagonal nanostructure of the original silica matrix is preserved and appears with clusters of apatite nanocrystals, which have dimensions of approximately 20 nm indicated with the arrows. The observations also revealed some distortion with respect to the regular original silica channels and dark features located in the channels of the silica matrix which suggest the growth of HA clusters inside the pores of the silica matrix. A non-uniform variation in contrast along the tunnels can be systematically observed. This might be related with the inclusion of HA material inside the channels. Fig. 2 shows a high magnification micrograph of HA nanocrystals showing lattice fringes separations of 3.09 and 3.44 which correspond to the (2 1 0) and (0 0 2) HA reflections. Peak width analysis of a slow-scanned (0 0 2) reflection of the diffraction pattern indicates an average crystal size of 19 nm which is in good agreement with dimensions measured using TEM.

X-ray powder diffraction analysis of the SBA-15 and the HA-SBA-15 materials are presented in Fig. 3. The small-angle X-ray pattern shows three resolved peaks with $d$ spacing of 98, 56 and 48 Å (9.8, 5.6 and 4.8 nm) for the SBA-15 and of 109, 62 and 53 Å (10.9, 6.2 and 5.3 nm) for HA-SBA-15 material which are indexed as the (1 0 0), (1 1 0) and (2 0 0) reflections of a two-dimensional hexagonal mesostructure with "$a$" lattice constant of 112 Å (11.2 nm) and 124 Å (12.4 nm) respectively in the space group $P6mm$. This result confirms the preservation of the nanostructure in HA-SBA-15 material which shows a similar pattern to the bare SBA-15 with a somewhat large lattice constant "$a$". The comparison of the relative intensities of the $d_{110}$ and $d_{100}$ values reveals likewise a good mesoscopic order for the HA-SBA-15 material. The wide-angle X-ray diffraction pattern of the HA-SBA-15 material is also shown in Fig. 3 (inset) and indicates the successful formation of HA crystals as a single crystalline phase in the material. Characteristic reflections from planes of the pure HA phase are indicated in Fig. 3 (inset). All the peaks matched well to those of HA phase (JCPDS 9-432), and no other peaks were detected.

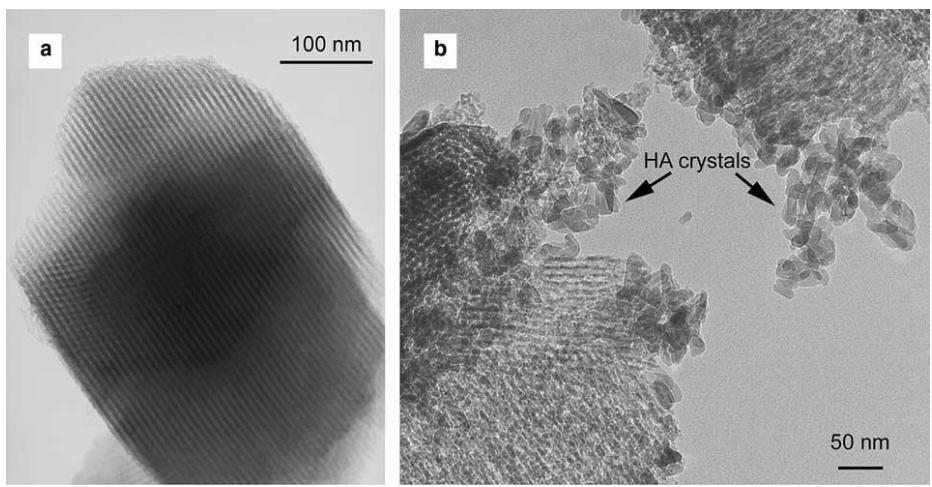

Fig. 1. TEM images showing the base SBA-15 silica structured matrix material (a) and a general low magnification image of the nanohydroxyapatite crystals within the structured silica host matrix in the HA-SBA-15 material (b).

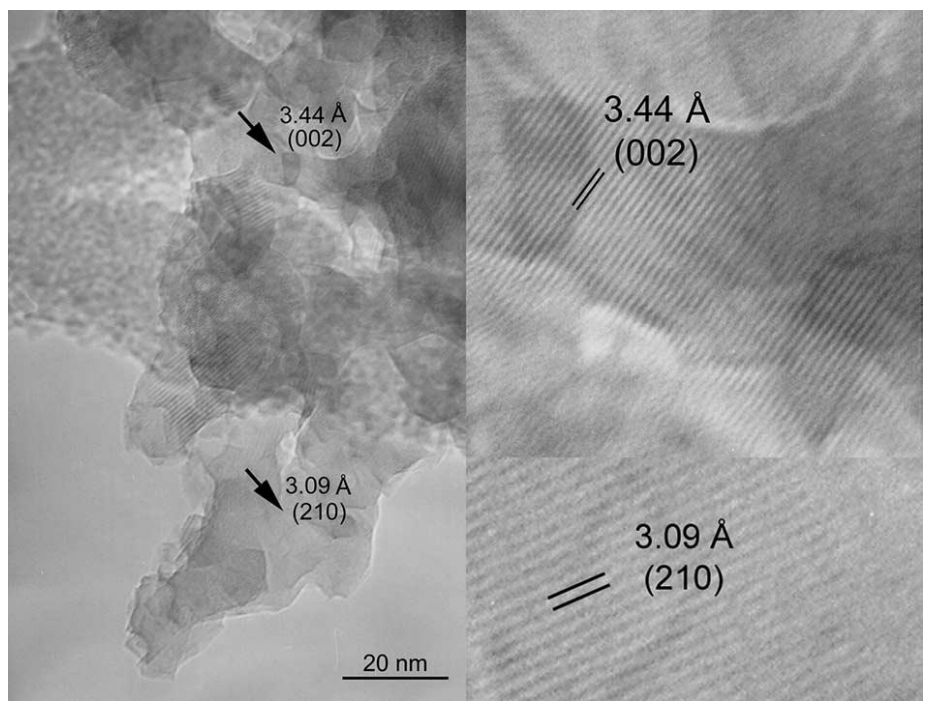

Fig. 2. High magnification image corresponding to the HA nanocrystals showing 3.09 and 3.44 lattice fringe separations in the HA-SBA-15 material.

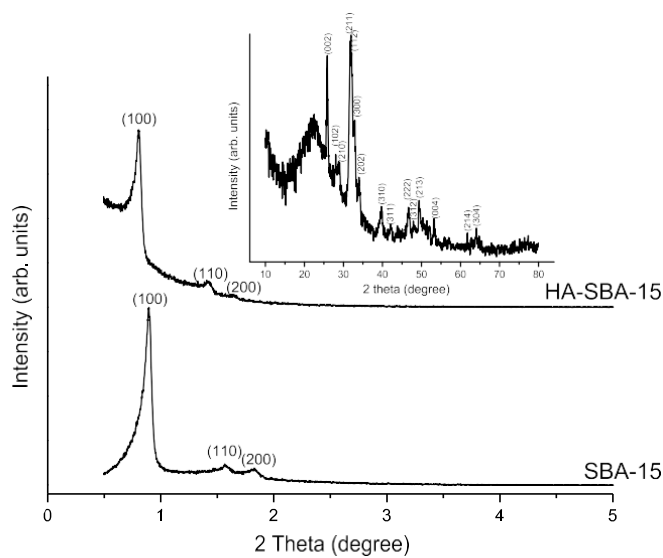

Fig. 3. SXRD patterns of the SBA-15 and HA-SBA-15 materials. WXRD pattern of the HA-SBA-15 material (inset).

Quantitative elemental analysis of the HA-SBA-15 material was carried out using X-ray fluorescence. The Ca/P ratio was found to be 1.9 which is close to the 1.7 ratio expected for the HA stoichiometry. Nevertheless, percentages below the XRD detection limits of any other calcium phosphate stoichiometry different to HA, calcium silicate or other calcium species can not be completely disregarded. Table 1 shows the elemental composition, textural and structural parameters.

An assessment of the porosity of the reference SBA-15 and the HA-SBA-15 material has been carried out using $N_2$ adsorption/desorption isotherms. The comparison of HA-SBA-15 with the SBA-15 specific surface area shows a dramatic decrease. Both samples exhibit a type IV isotherm characteristic of mesoporous materials with microporous contributions. SBA-15 material shows the type-H1 hysteresis loop characteristic of mesoporous materials with one-dimensional cylindrical channels. HA-SBA-15 shows in contrast a H3-type loop which is often observed with aggregates of plate-like particles that give rise to slit-shaped pores. $N_2$ adsorption/desorption isotherms are presented in Fig. 4. BJH desorption pore distributions for SBA-15 and HA-SBA-15 are presented in Fig. 5. SBA-15 material shows a narrow pore size distribution corresponding to 50 Å (5.0 nm). HA-SBA-15 shows a broader distribution with diameters within 70–500 Å (7.0–50.0 nm). The results indicate that HA growing crystals block completely the 50 Å (5.0 nm) pore size while increasing the average pore size diameter.

Table 1
Physicochemical characterization

| Material | $N_2$ adsorption–desorption | | | XRD | XRF | $^{29}$Si NMR |
|---|---|---|---|---|---|---|
| | Surface area, m$^2$ g$^{-1}$ | Pore diameter, Å | Pore volume, cm$^3$ g$^{-1}$ | $d_{100}$,[a] Å | Elemental atomic ratio | Q$^3$/Q$^4$ ratio |
| SBA-15 | 761 | 50 | 0.92 | 98 | | 0.6 |
| HA-SBA-15 | 273 | 140 | 1.08 | 109 | Ca:P = 1.9; Ca:Si = 0.37 | 0.1 |

[a] $d$-Values of (100) reflection.

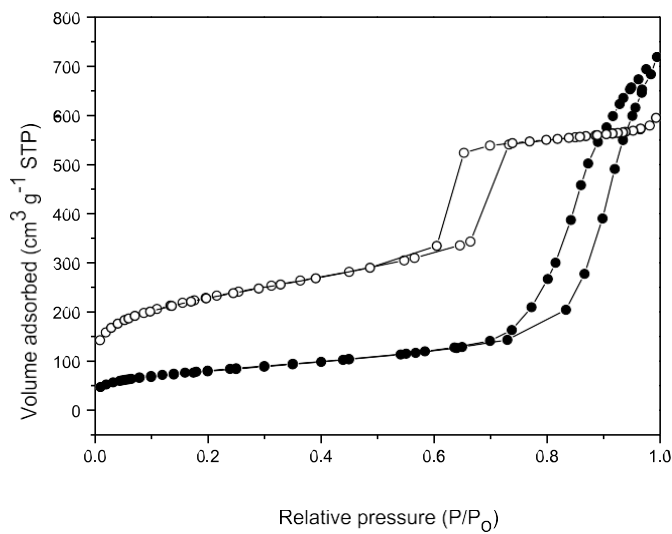

Fig. 4. Nitrogen adsorption–desorption isotherms: (s) SBA-15 material; (d) HA-SBA-15 material.

A comparison of the ''$a$'' unit cell parameter of the $P6mm$ structure for the SBA-15 and HA-SBA-15 materials indicates that the parameter remains very similar in both materials. The decrease in surface area and the elimination of the 50 Å (5.0 nm) diameter pores shown in Fig. 5, while retaining the ''$a$'' unit cell parameter, suggest that the original mesoporous are filled by HA crystals.

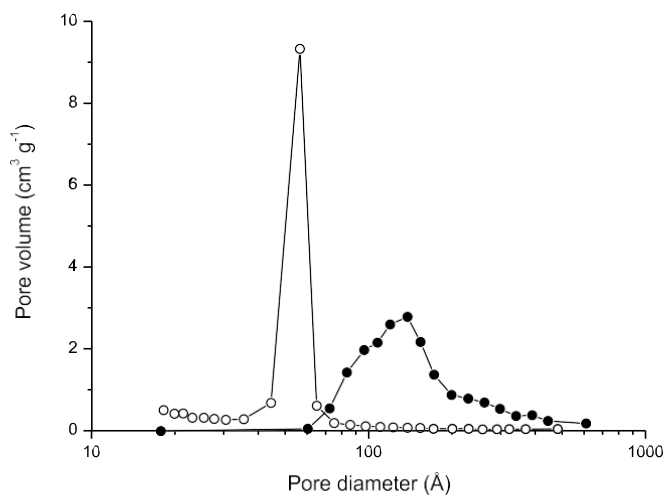

Fig. 5. Pore size distributions. (s) SBA-15 material; (d) HA-SBA-15 material.

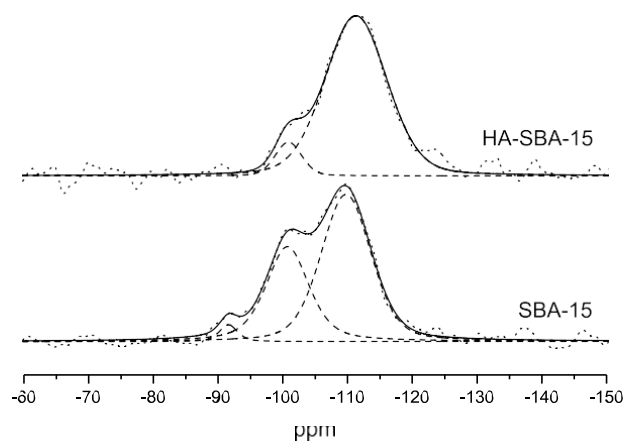

Fig. 6. $^{29}$Si MAS NMR spectrum of materials. Dotted line: experimental spectrum; dashed lines: deconvoluted peaks; solid line: calculated spectrum.

The pores within the mesopore range have been willing to act as initiation sites for HA crystal nucleation. Supersaturation of surface potentials inside the pores which produce increasing levels of calcium and phosphate ions will more likely allow the precipitation of HA inside the pores.

$^{29}$Si MAS-NMR techniques have monitored the changes of the silanol group concentrations. HA-SBA-15 and SBA-15 spectra are shown in Fig. 6. SBA-15 solid spectra clearly exhibit three peaks associated with $Q^2$ (ca. −90 ppm), $Q^3$ (ca. −100 ppm), and $Q^4$ (ca. −110 ppm) Si connectivities [17,18]. HA-SBA-15 shows only peaks $Q^3$ and $Q^4$. Peaks in the $Q^2$ and $Q^3$ region are due to silanol groups, while those in the $Q^4$ region are attributed to Si connected to four tetrahedral atoms through O atoms. Therefore, the ratio of $Q^3/Q^4$ signals obtained by deconvolution of the overall peak provides a measurement of the amounts of silanol groups present in the samples [19]. Deconvolution of the experimental NMR spectra has been fitted using the dmfit Program [20] and are also shown in Fig. 6. $Q^3/Q^4$ ratios for the SBA-15 and the HA-SBA-15 are presented in Table 1. Ratio value comparisons show a strong reduction of silanol groups for the HA-SBA-15 material. There is also a slightly shift to lower frequencies and a signal broadening of the line of $Q^4$ groups for the HA–silica composite material. Changes in Si–O bond distances and Si–O–Si angles in sil- icates have been shown to contribute to signal broadening and cause a shift of the line of $Q^4$ groups [21].

The growth of HA in the mesoporous silica has been achieved by means of a synthesis route which includes a high pH step of ∼9 when phosphate ions were incorpo- rated. Rates of hydrolysis, condensation and redissolution of TEOS produced by pH changes indicates that when pH > 4 the condensation rate is not only proportional to the concentration of OH$^-$ anions but also superior to the hydrolysis rate [22]. The pH increase produces condensation of adjacent silanol groups which yields new oxo bridges and consequently an important reduction of the silanol (Si–(OSi)$_3$–OH) groups. Our hypothesis is that the siloxane cavities formed in this way might retain calcium metal ions stabilizing them in a coordination favored by some kind of macrocyclic effect. This coordination may provide sites for the nucleation of HA crystals with subsequent pore filling of the silica template.

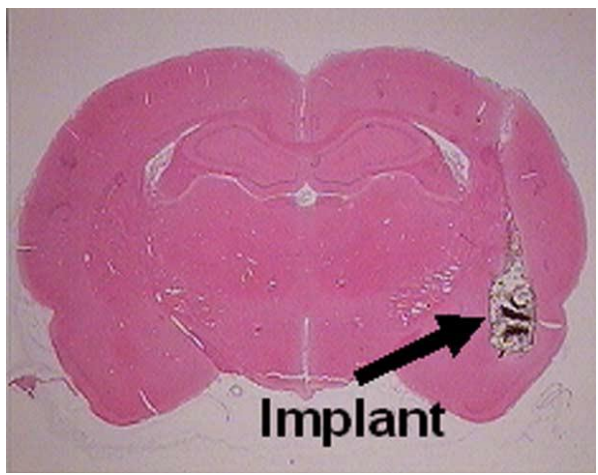

Fig. 7. SBA-15 implant material in the temporal lobe of the rat.

The local effect of the silica SBA-15 material implant has been studied in the close vicinity, with the brain tissue nearby. The SBA-15 material cylinder was surgically implanted in the adult male rats as has been described in Section 2. A well-organized fibrous capsule was formed around the implant at 14 d and later. Fig. 7 presents the SBA-15 implant in the temporal lobe of the rat. No pathology was reported. The implant did not cause necrosis or inflammation. An optical micrograph of the implant and the tissue nearby after 120 days of implantation is shown in Fig. 8. The figure shows that the surrounding biological material has been self-readapted to the contour of the inorganic material. The boundary of the implant displayed no gliosis. The corresponding neurons of this zone show no major affection, demonstrating a well adapted biological brain region. The results show a good biocompatibility for brain tissue and open the potential application to other kind of tissues. The central nervous system is considered an organ with a particular vulnerability and immunological status [23,24]. The brain biocompatibility can be therefore seen as a good test for the preliminary study of the biocompatibility of the SBA-15 materials. Brain tissue biocompatible tests using the HA-SBA-15 material are currently under way. The first experiments which are not shown in the manuscript indicate no pathology after implantation. A detailed report showing HA-SBA-15 biocompatibility studies will be the subject of a future paper. Further research will concentrate on studying material–bone tissue specific responses and its potential application as bone scaffold substitute.

The high surface area, macroscopic composition and nano dimensions of the HA crystals in HA-SBA-15 highlights this material's potential for use as a bone tissue scaffold biomaterial. Additionally, its textural properties make this a very interesting material for its application in the local delivery of chemotherapy drugs after surgery [25].

## 4. Conclusions

The synthesis of a novel biomaterial consisting of hydroxyapatite nanocrystals grown within a mesostructured silica matrix is reported. The 2D-hexagonal nanostructure of the original silica template is preserved and is evident with clusters of apatite nanocrystals of approximately 20 nm. The growth of HA nanocrystals inside the mesopores of the silica matrix was also reported. $^{29}$Si MAS-NMR analysis has shown a strong reduction of the silanol groups concentration with HA growing within the base SBA-15 material.

In vivo studies using the base matrix implant material SBA-15 are reported here for the first time. The results show excellent biocompatibility with brain tissue. Biological studies and brain tissue biocompatibility tests using the HA-SBA-15 material are currently under way.

## Acknowledgements

We gratefully acknowledge the financial support provided by the DGICYT (Project MAT2003-06540-C02-01). A. D´ıaz thanks the ''Ramón y Cajal Programme'' (Project no. 2003/229).

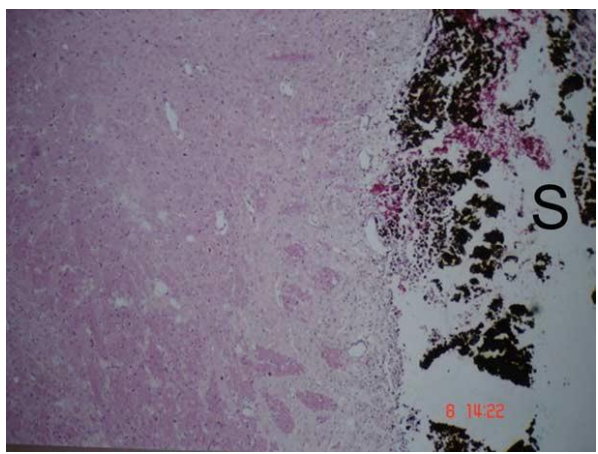

Fig. 8. Optical micrograph of the SBA-15 implant (S) and the tissue nearby in a rat brain after 120 days of implantation.